\documentclass[11pt]{article}
\usepackage{moriond,epsfig}

\def\Journal#1#2#3#4{{#1} {\bf #2} (#4) #3}

\def\PLB{{\em Phys. Lett.}  B}
\def\PRL{{\em Phys. Rev. Lett.}}

\def\JHEP{{\em J. High Energy Phys.}}
\def\EPJC{{\em Eur. Phys. J.} C}

\def\be{\begin{equation}}
\def\ee{\end{equation}}
\def\bea{\begin{eqnarray}}
\def\eea{\end{eqnarray}}

\newcommand{\met}{\ensuremath{\not\!\!\! E_{\mathrm{T}}}}
\newcommand{\fbinv}{\,{\rm fb}^{-1}}

\begin{document}
\vspace*{4cm}
\title{SUPERSYMMETRY SEARCHES WITH ATLAS AND CMS}

\author{Steven Lowette}

\address{University of California, Santa Barbara, USA\\[2mm] \emph{on behalf of the ATLAS and CMS collaborations}}

\maketitle\abstracts{Highlights are presented of current searches for supersymmetry produced in proton-proton collisions at the LHC, operating at $\sqrt{s} = 7 \,{\rm TeV}$. Searches with missing transverse energy are summarized from both the ATLAS and CMS experiments. A wide range of final states is covered, including recent searches with photons, leptons, and purely hadronic final states. Future developments are anticipated with extra focus on exclusive searches in electroweak and third-generation supersymmetry production.}

\section{Searching for supersymmetry at the LHC}

Supersymmetry~\cite{susy} (SUSY) postulates a new symmetry, linking bosons and fermions. As such, SUSY brings an extension to the Standard Model (SM) of particle physics, in which a whole spectrum of new particles is predicted. Indeed, the non-observation of supersymmetric partners thus far requires supersymmetry to be a broken symmetry, and several breaking mechanisms have been proposed, like gravity mediation, gauge mediation, or anomaly mediation, each giving rise to models with specific characteristic phenomenological properties.

Supersymmetry is appealing, and remains to date, because it potentially solves several problems of the Standard Model. SUSY can provide an elegant solution to the apparent need for fine-tuning in the calculation of the Higgs boson mass; it leads to a unification of the gauge coupling strengths at a common scale; and when imposing so-called $R$ parity to restore proton stability, the lightest supersymmetric particle (LSP) becomes stable, providing an excellent dark-matter candidate.

Supersymmetry has been searched for at the Large Hadron Collider (LHC) in proton-proton collisions at $7 \,{\rm TeV}$ centre-of-mass energy, from  2010 onwards. At first, only squark and gluino pair production provided viable SUSY production search modes, with the coloured nature of these particles resulting in a sizeable production rate early on. Because of the complex decay phenomenology, driven by the large diversity in possible SUSY mass spectra, the searches were typically designed not to focus on particular models, but rather to look for deviations from SM background expectations for a diverse set of experimental signatures.

With the advent of more than hundred times larger data samples in 2011, production modes with lower cross sections have become accessible. This opens up the possibility for more dedicated searches in exclusive channels, like electroweak chargino and neutralino production, or direct stop or sbottom production.

In these conference proceedings, we summarize a few highlights of the searches for supersymmetry as they were at the moment of the conference, at both the ATLAS~\cite{atlas} and CMS~\cite{cms} detectors. All searches reported here look for signatures including missing transverse momentum (\met), induced by the LSP escaping detection. Non-\met-based SUSY searches were presented elsewhere~\cite{DavidAdams,MortenJoergensen}, and third-generation SUSY results from ATLAS can be found in a dedicated contribution~\cite{AntoineMarzin}.

More specifically, we report on searches with photons from both experiments~\cite{CMS-PAS-SUS-12-001,arXiv_1111_4116}, an ATLAS search for disappearing tracks~\cite{arXiv_1202_4847}, a glimpse of hadronic analyses for both CMS~\cite{CMS-PAS-SUS-12-005} and ATLAS~\cite{arXiv_1110_2299}, inclusive searches with same-charge (same-sign) lepton pairs~\cite{SUS-11-010-5fb,ATLAS-CONF-2012-004}, searches from both experiments with three or more leptons~\cite{CMS-PAS-SUS-11-013,ATLAS-CONF-2012-023}, and a CMS analysis looking for same-sign dileptons in the presence of ${\rm b}$-tagged jets~\cite{CMS-PAS-SUS-11-020}.
This overview only attempts to discusses a selection of results, and since then many more have become available. The complete and up-to-date lists of supersymmetry results from both experiments can be found here~\cite{ATLASpublic,CMSpublic}.


\section{Searches with photons}

A search for supersymmetry in final states with photons was pursued in CMS with two selections. One selection requires a single photon to be accompanied by two jets, with $\met > 100 \, {\rm GeV}$; a second selection asks for two photons, in this case relaxing the other requirements by asking a single jet and $\met > 50 \, {\rm GeV}$. The former analysis was performed with an integrated luminosity of $4.3\fbinv$, while the latter used $4.7\fbinv$ of data. The ATLAS experiment pursued a similar search with $1.07 \fbinv$ of data, in this case using a two-photon selection, requiring one additional jet and $\met > 100 {\rm GeV}$.

Standard Model backgrounds for all three analysis are mainly arising from QCD multijet production, with real photons or jets faking photons, and from electroweak processes in the form of electrons faking photons. The multijet background is determined by extracting the \met{} distribution from an appropriate control sample, where the ATLAS analysis in addition cross-checked their prediction by modelling the \met{} spectrum with a ${\rm Z \to e^+ e^-}$ control sample. Electron backgrounds are predicted using ${\rm e} \to \gamma$ misidentification rates measured in data, while other smaller backgrounds are estimated from simulation.

\begin{figure}[htb]
  \begin{center}
    \raisebox{-0.5\height}{\includegraphics[width=0.32\textwidth]{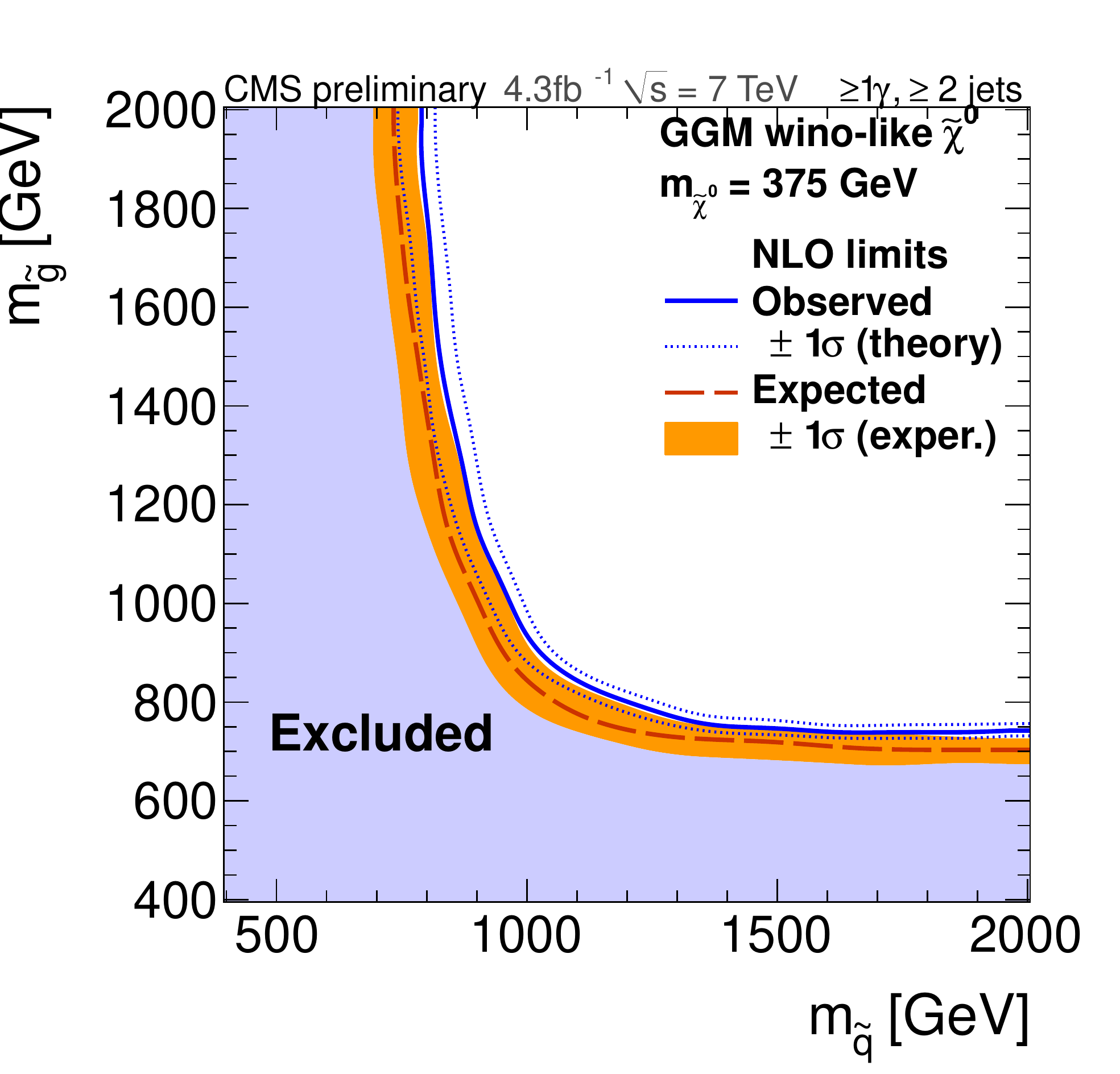}}
    \raisebox{-0.5\height}{\includegraphics[width=0.32\textwidth]{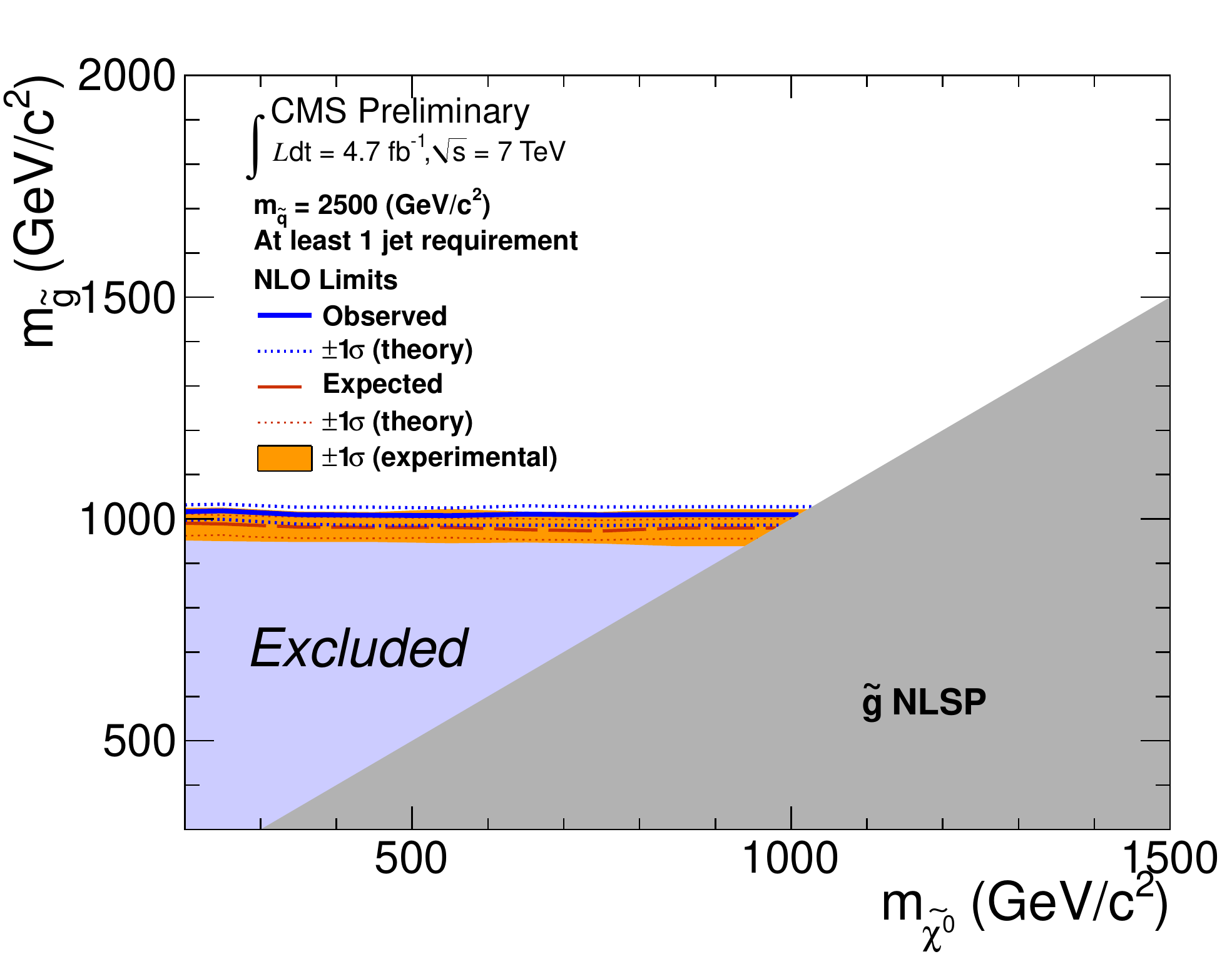}}
    \raisebox{-0.5\height}{\includegraphics[width=0.32\textwidth]{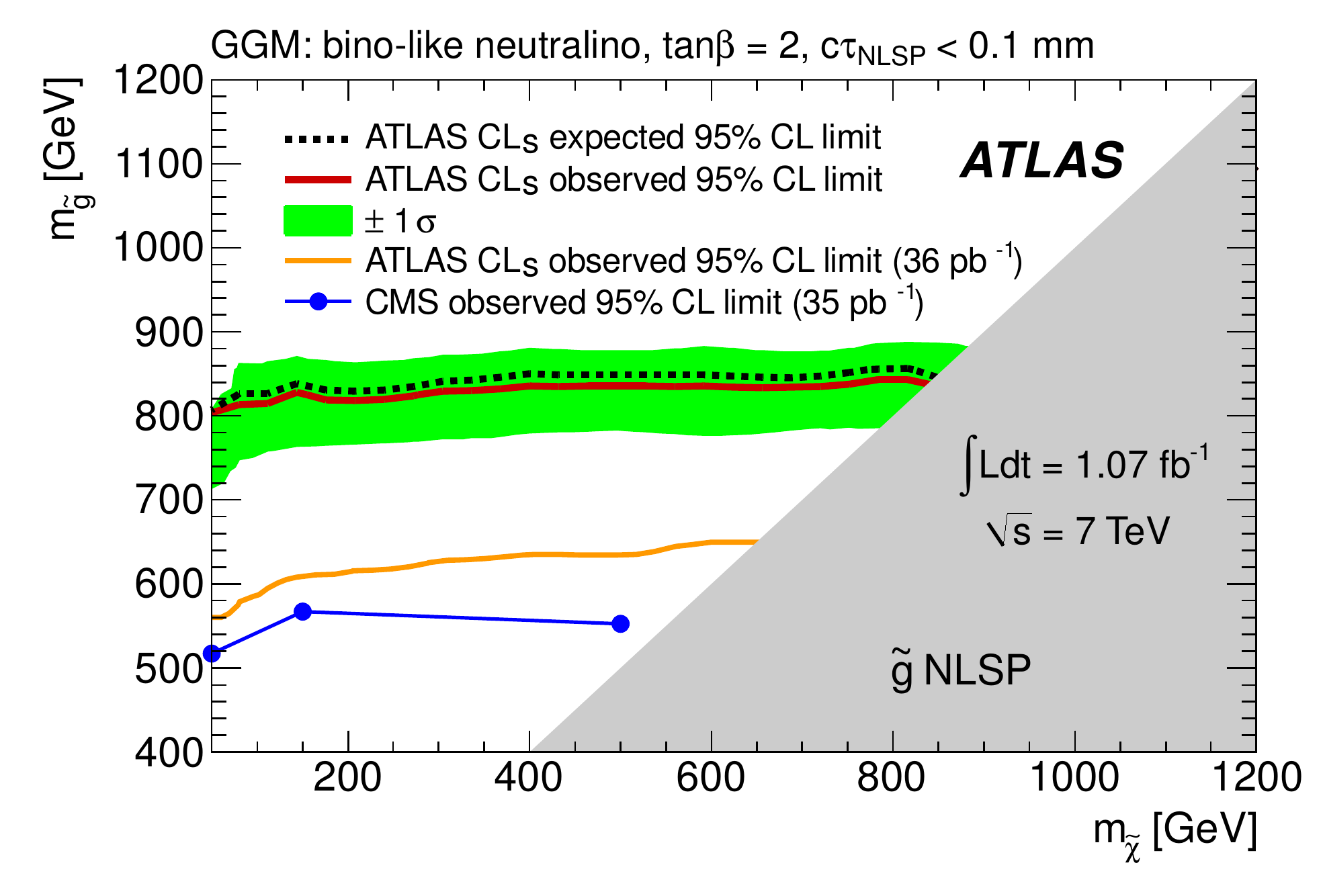}}
  \end{center}
  \caption{Exclusion limits at 95\% CL for the single photon (left) and diphoton selection (middle) for the CMS analyses, and for the diphoton search from ATLAS (right).}
  \label{fig:photons}
\end{figure}

Signs of new physics were sought by looking for an enhancement of the tail of the \met{} distributions. The data was found to be in good agreement with the predicted \met{} spectra. Interpretations of the results were made within a general gauge mediation context, with the gravitino as low-mass LSP. In such scenarios the phenomenology of the models is driven by the next-to-lightest supersymmetric particle (NLSP). Simplified models were considered with squark and gluino production, followed by decay through a bino-like neutralino NLSP (one-photon selection) and a wino-like neutralino NLSP (two-photon selections). In Figure~\ref{fig:photons} the resulting parameter space excluded at 95\% confidence level (CL) is shown for the different analyses, as a function of squark, gluino, and NLSP mass. The limits are found to be mainly driven by the production cross section, with little impact from the neutralino mass.

\section{Search for disappearing tracks}

The ATLAS experiment performed a dedicated search for disappearing tracks. The phenomenology is motivated by an anomaly-mediation scenario (AMSB), where the first chargino $\tilde{\chi}_1^\pm$ and neutralino $\tilde{\chi}_1^0$ differ in mass by only about $200 \,{\rm MeV}$. This results in the chargino possibly being observed travelling a significant distance within the tracker volume before decaying to the neutralino, leading to \met{}, a high-momentum track that apparently ``disappeared'', and a soft pion. Charginos are considered here to be produced in cascade decays from gluinos and squarks. The selection therefore requires the presence of three jets, accompanied by $\met > 130\,{\rm GeV}$, and a track with less than 5 hits in the outer part of the transition radiation tracker (TRT), the outermost layer of the ATLAS inner tracking system.

\begin{figure}[htb]
  \begin{center}
    \raisebox{-0.5\height}{\includegraphics[width=0.48\textwidth]{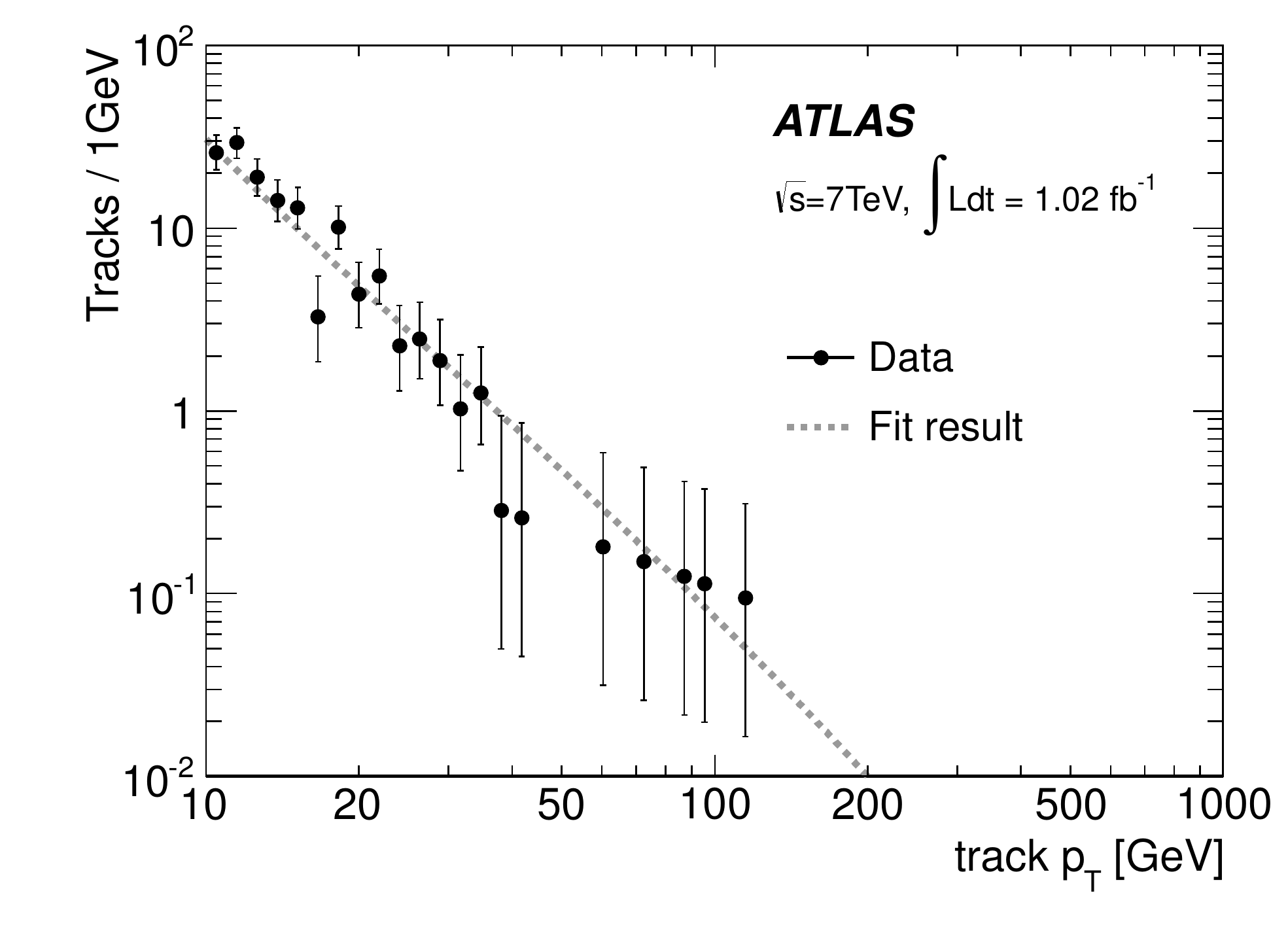}}
    \raisebox{-0.5\height}{\includegraphics[width=0.48\textwidth]{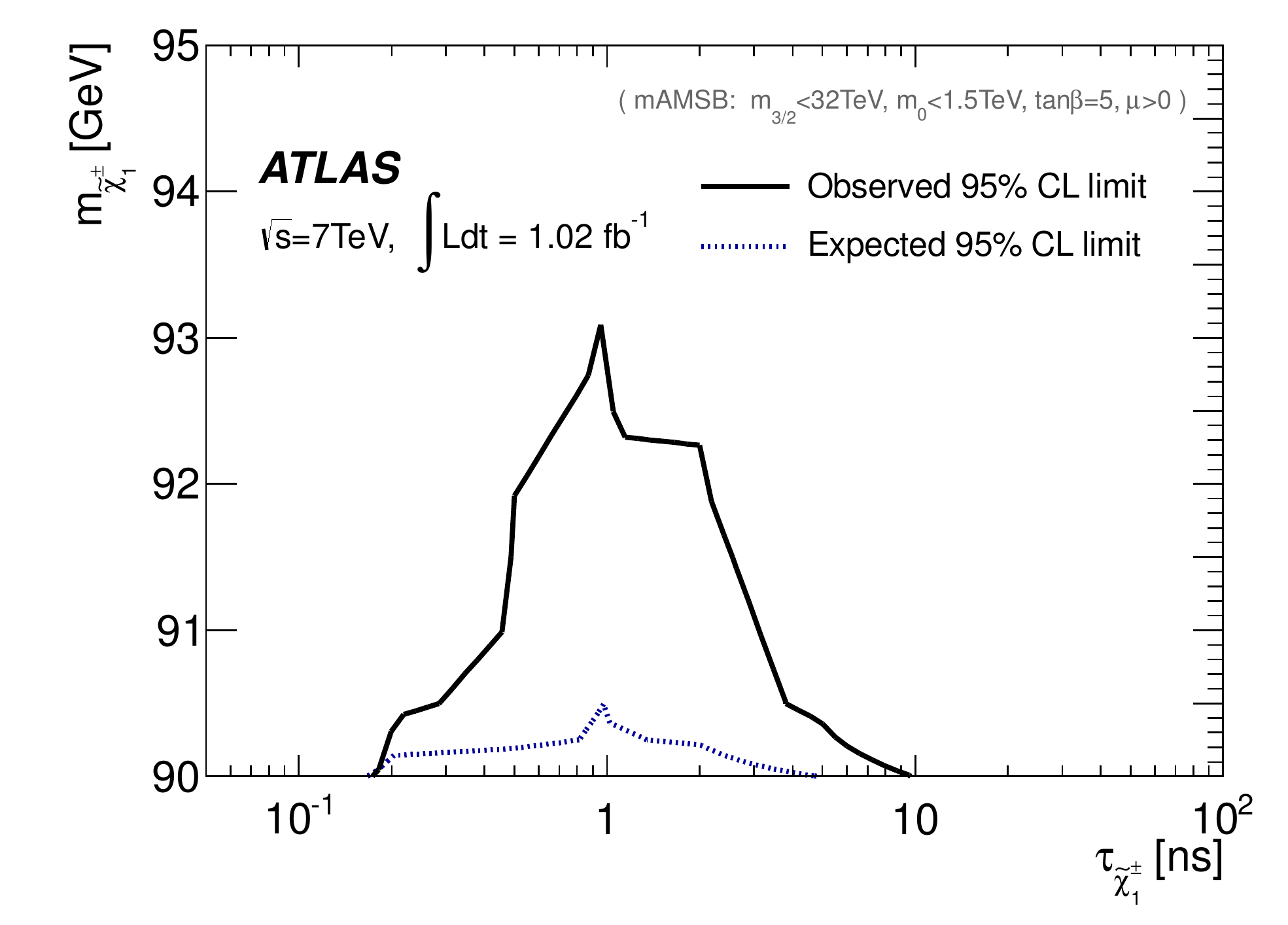}}
  \end{center}
  \caption{Spectrum of the ``disappearing'' tracks in data, on top of the prediction from data (left); 95\% CL limits in the considered AMSB scenario as a function of the chargino mass and lifetime (right).}
  \label{fig:disapptracks}
\end{figure}

Backgrounds to this signature arise from high-momentum tracks interacting with the TRT material, producing at the point of interaction a non-reconstructed kink or shower, or from low-momentum charged particles with a bad momentum measurement, due to large multiple scattering. An estimation of these backgrounds from data was developed, applying a combined shape fit with components determined in appropriate control samples. The momentum spectrum of ``disappearing'' tracks is shown for data and the background prediction in Figure~\ref{fig:disapptracks} (left), for a data sample of $1.02 \fbinv$. In Figure~\ref{fig:disapptracks} (right) an interpretation in an AMSB model is made (with parameters as given in the figure), showing 95\% CL exclusion limits as a function of the chargino mass and lifetime.

\section{Inclusive and multijet searches}

On the side of the hadronic searches for squark and gluino production, an inclusive analysis was performed with the so-called ``razor'' variables $R$ and $M_R$, using $4.4 \fbinv$ of integrated luminosity. The analysis is actually sensitive beyond hadronic searches, because it combines exclusive single-lepton and dilepton data samples in a global fit. The razor $R$, sensitive to the ratio of missing and visible momentum in the event, is used to suppress backgrounds, in particular QCD multijet events. $M_R$, on the ohter hand, peaks broadly at the characteristic scale of the physics process. The tails of both the $R$ and $M_R$ distributions can be described excellently with a two-component exponential for all backgrounds.

\begin{figure}[htb]
  \begin{center}
    \raisebox{-0.5\height}{\includegraphics[trim=10mm 25mm 0cm 0cm,clip=true,width=0.48\textwidth]{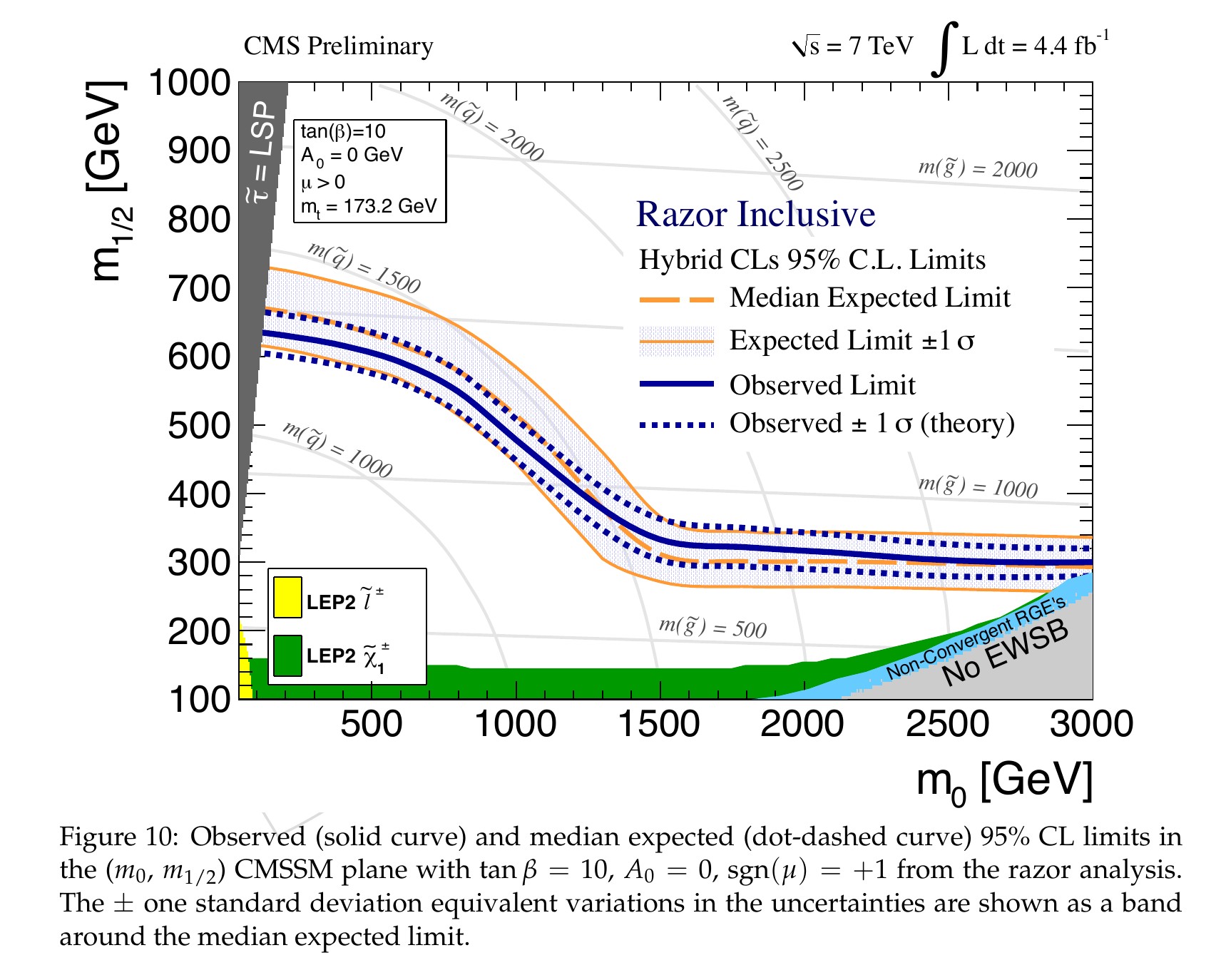}}
    \raisebox{-0.5\height}{\includegraphics[width=0.48\textwidth]{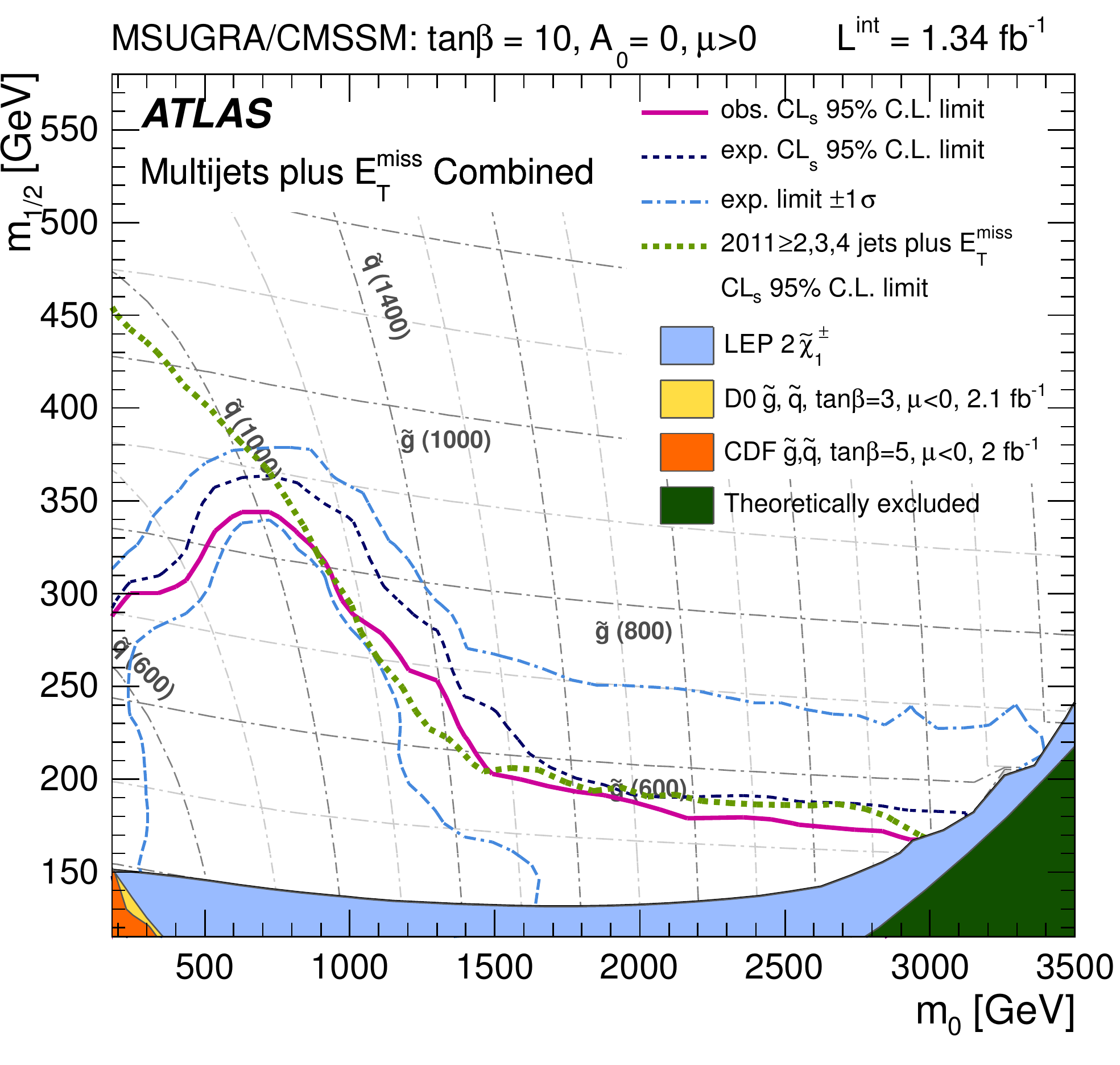}}
  \end{center}
  \caption{Interpretation of the CMS inlusive razor search (left) and the ATLAS multijets analysis (right) in the CMSSM $(m_0 , m_{1/2})$ plain.}
  \label{fig:hadronic}
\end{figure}

The background strategy consists of fitting such a background description in both $R$ and $M_R$ dimensions, simultaneously using samples with zero, one, and two leptons. This function is then extrapolated into the signal region. The data is found to be in good agreement with the background description, and limits on new physics were derived in the Constrained Minimal Supersymmetric Standard Model (CMSSM), as shown in Figure~\ref{fig:hadronic} (left) as a function of the model parameters $m_0$ and $m_{1/2}$.

While most hadronic analyses, like the ``razor'', are mostly sensitive to low jet multiplicities, longer decay chains can give rise to multijet final states.
The ATLAS experiment performed a dedicated hadronic analysis with between 6 and 8 jets in the events, using $1.34 \fbinv$ of data. The search variable used here, $\met / \sqrt{H_{\rm T}}$, with $H_{\rm T}$ the scalar sum of the jet momenta, is a measure of the significance of the \met{} in the event.

The particular challenge for this analysis is the presence of a large number of events from QCD multijet production. This background is estimated from data in control samples with a lower jet multiplicity, and is normalized to the data in a background-dominated region of low \met{} significance. Data was found to be in good agreement with the background predictions, and in Figure~\ref{fig:hadronic} (right) this absence of signal is translated into a limit in the CMSSM parameter space. The focus on the high jet multiplicity is found to increase sensitivity in particular at intermediate and high $m_0$.

\section{Search with same-sign dileptons}

Searches for supersymmetry with leptons make it possible to evade the large backgrounds from multijet production, while keeping the \met{} requirements modest. This is in particular the case for the search for same-charge (same-sign) leptons, a signature which can arise in decays from gluino pair production as well as same- or opposite-sign squark production. In addition, the use of low-momentum leptons makes it possible to more easily probe compressed SUSY spectra, where the LSP is close in mass to heavier states, and little energy from the SUSY decays remains observable in the event.

Both CMS and ATLAS carried out searches for same-sign dileptons. In the case of CMS an integrated luminosity of $4.7\fbinv$ was analysed in several search regions, each probing different mass splitting scenarios (high and low $H_{\rm T}$ and \met, low-momentum leptons), including tau leptons. The ATLAS collaboration analysed $2.05\fbinv$ of data, in a search region with electron or muon same-sign pairs, four jets and $\met >150\,{\rm GeV}$.

The main analysis challenge comes from the need to estimate the background of non-prompt leptons arising from jets, and the one from lepton charge mis-identification. Multiple methods were developed to measure both backgrounds from data, confirming the charge mis-identification to be very small. Apart from the non-prompt leptons, a second background component contributes from rare processes giving rise to final states with prompt leptons, like diboson (including same-sign ${\rm WW}$) and ${\rm t\bar{t} W/Z}$ production. These processes with low cross sections are gaining importance with the growing data samples. They are estimated from simulation.

\begin{figure}[htb]
  \begin{center}
\raisebox{-0.5\height}{\includegraphics[width=0.48\textwidth]{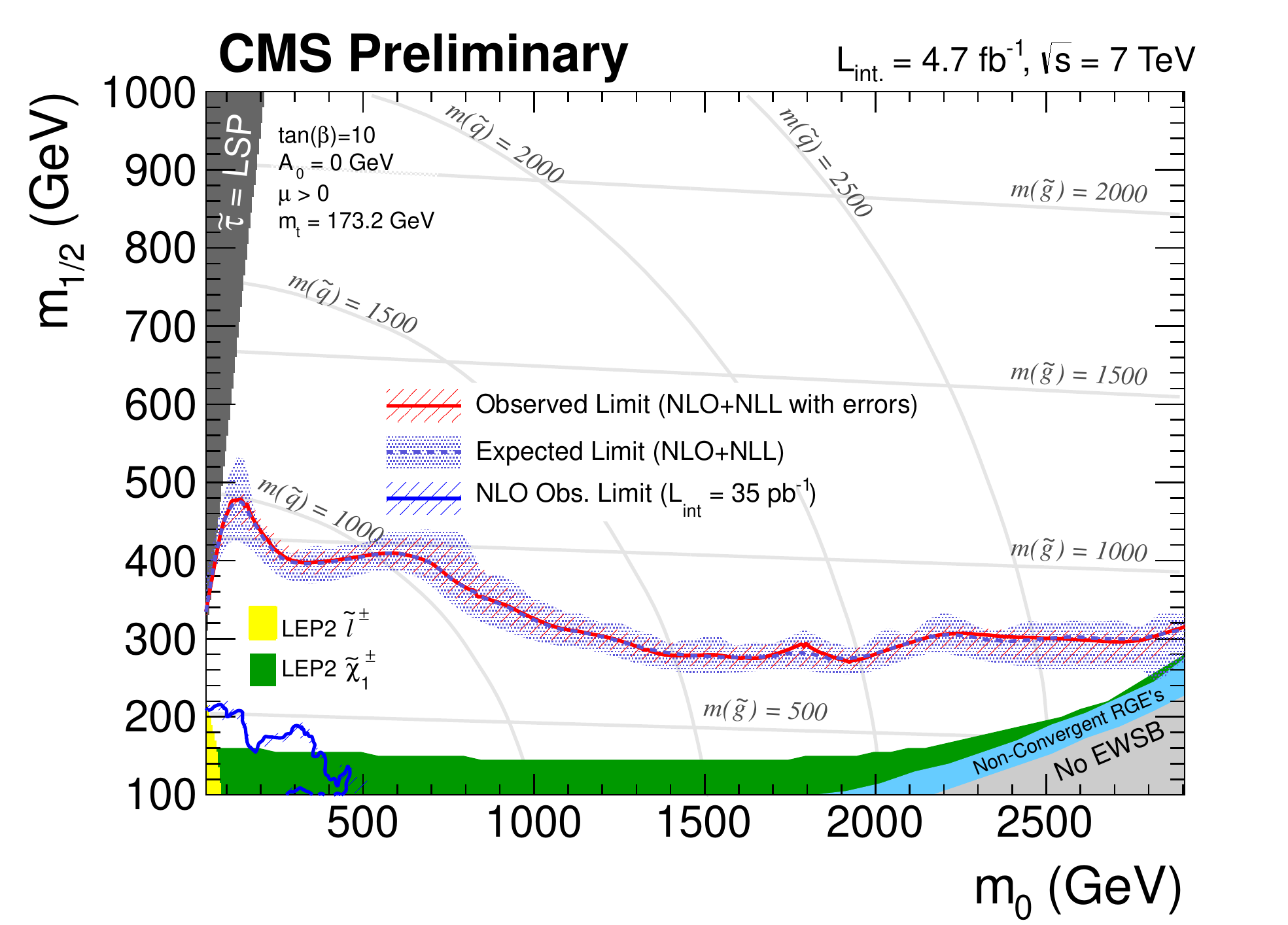}}
    \raisebox{-0.5\height}{\includegraphics[width=0.48\textwidth]{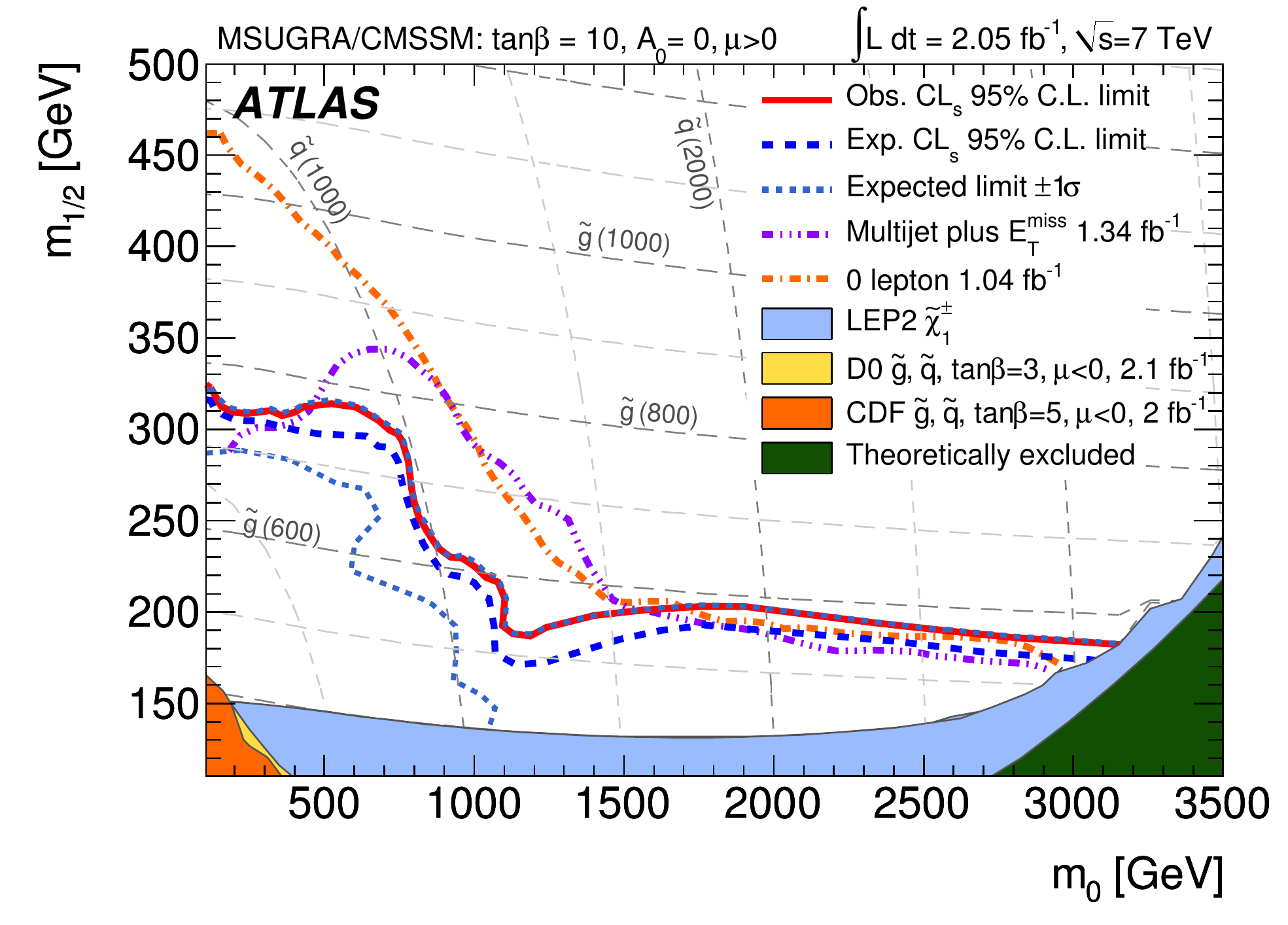}}
      \end{center}
  \caption{Interpretation of the same-sign CMS search (left) and ATLAS search (right) in the CMSSM $(m_0 , m_{1/2})$ plain.}
  \label{fig:ss}
\end{figure}

Across all search regions, the data is found to be in good agreement with the expected background, and limits on new physics have been obtained within a variety of models with same-sign dilepton production. The exclusion obtained in a projection of the CMSSM parameter space is shown in Figure~\ref{fig:ss}.

\section{Multilepton searches}

When searching for supersymmetry with three or four leptons, the number of possible search regions grows large. In CMS an analysis was performed with $4.7\fbinv$ of data, searching in exclusive bins of electron, muon or tau lepton multiplicity, comparing small versus large $H_{\rm T}$ and \met{}, the presence of an opposite-sign same-flavour lepton pair or not, and such a lepton pair falling in our out of a window around the ${\rm Z}$-boson mass.

The backgrounds to this search depend on the selection bin considered, but analogies can be drawn with the same-sign dilepton search: ${\rm Z}$+jets and ${\rm t \bar{t}}$ contribute when an additional non-prompt lepton is present, while rare diboson and ${\rm t\bar{t} W/Z}$ production provide backgrounds of prompt leptons. In addition, ${\rm Z}$-boson production contributes in another way as well, when leptonic decay products radiate a virutal or real photon, which then in turn converts respectively ``internally'' or ``externally''. The former case provides an equal abundance of electron and muon conversions, in particular asymmetric ones, while the latter creates only electron-positron pairs within material, rarely contributing because of the displacement of the resulting tracks.

\begin{figure}[htb]
  \begin{center}
    \raisebox{-0.5\height}{\includegraphics[width=0.48\textwidth]{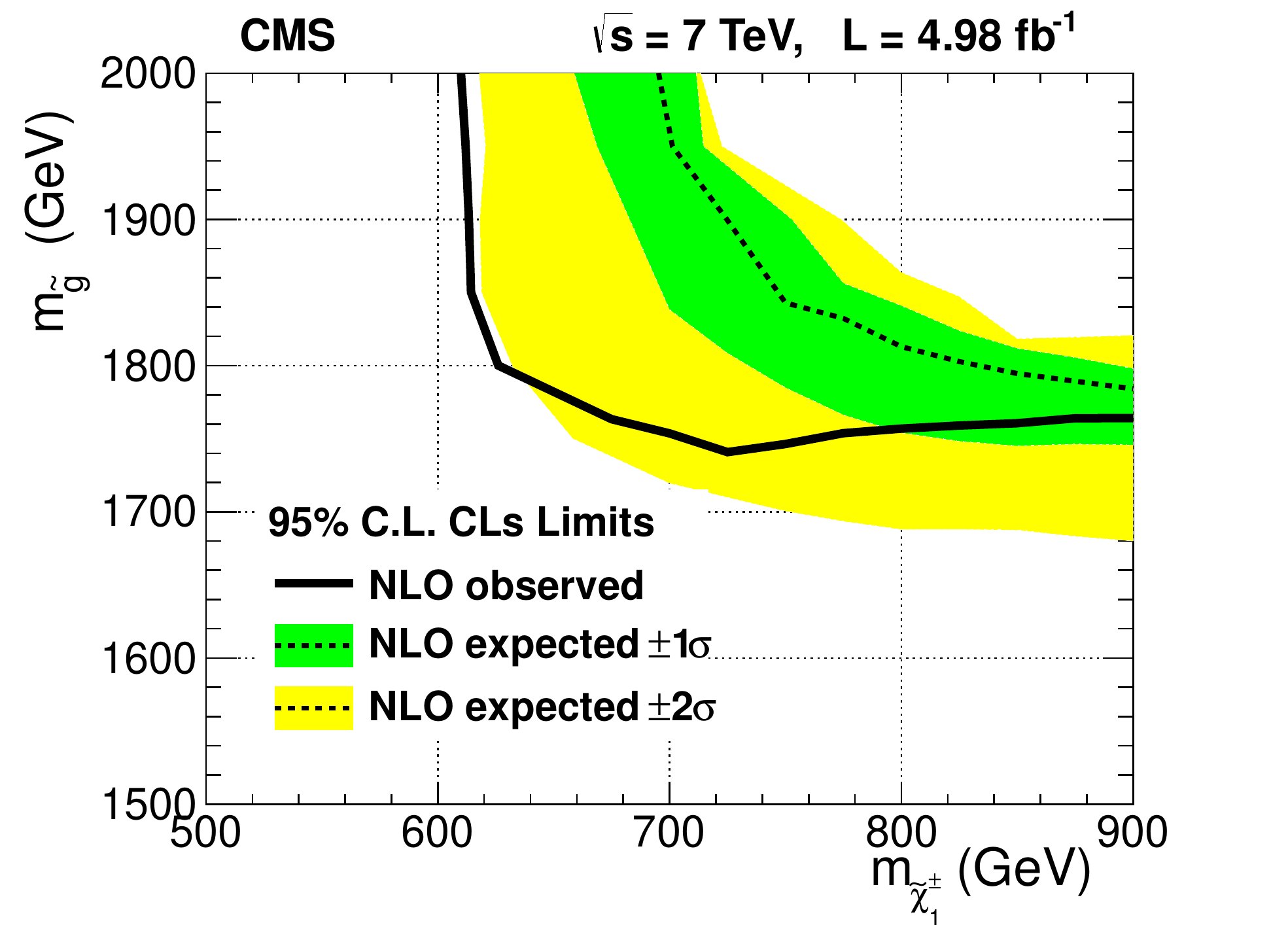}}
    \raisebox{-0.5\height}{\includegraphics[width=0.48\textwidth]{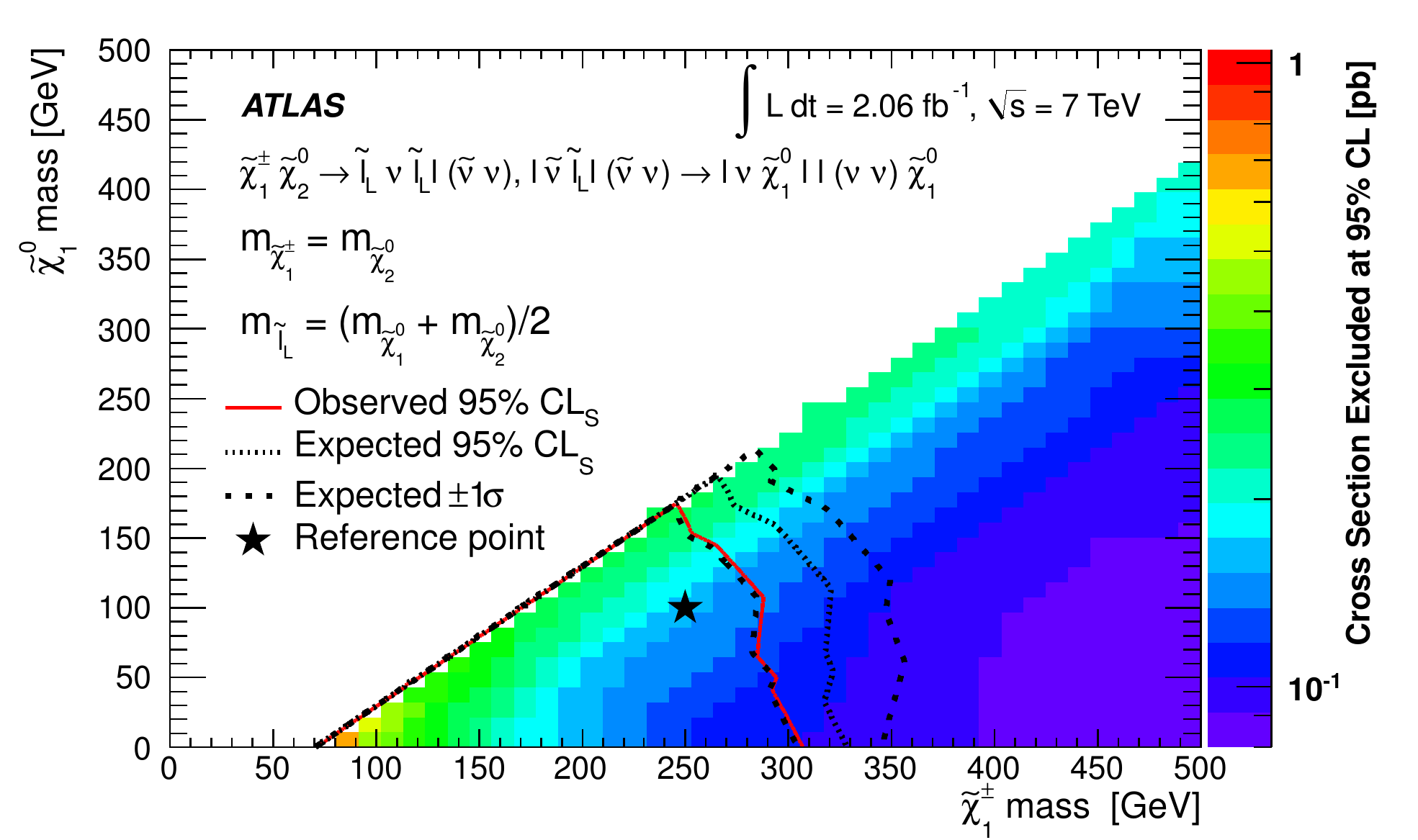}}
  \end{center}
  \caption{Exclusion reach for the CMS (left) and ATLAS (right) multilepton searches, interpreted in simplified models with, respectively, a gauge-mediation scenario and electroweak chargino-neutralino production.}
  \label{fig:multilep}
\end{figure}

The analysis' multi-bin approach provides simultaneous sensitivity to many possible new-physics scenarios, with both strong and electroweak production. One interpretation was carried out in a gauge-mediation inspired simplified model, with degenerate sleptons as NLSP, such that multilepton final states are obtained from decays of bino-like neutralinos through the sleptons to the gravitino LSP. In Figure~\ref{fig:multilep} (left) the 95\% CL limit for this model is shown, obtained from a combined fit over all selection bins. The limit is driven by a few bins that happen to show some discrepancy between data and background expectation, yielding a worse observed than expected limit. Over all search bins, though, the data is in fair agreement with the expectation.\\

A recent ATLAS analysis took a different approach, with a selection of three electrons and/or muons, $\met > 50\,{\rm GeV}$ and in particular no requirement on hadronic activity (with a b-jet veto for events without ${\rm Z}$-boson candidate). This selection was devised to obtain sensitivity for the low-cross section electroweak chargino-neutralino production, as opposed to the strong gluino and squark production.

The backgrounds to consider are of similar nature as in the CMS analysis. The data was found to be in good agreement with the expected background, and the result was interpreted in a simplified model with only electroweak production yielding a three-lepton final state. The mass of the produced $\chi_1^{\pm}$ and $\chi_2^0$ are assumed the same, with slepton masses exactly in the middle between the mass of $\chi_1^{0}$ and $\chi_2^0$. In Figure~\ref{fig:multilep} (right) the 95\% CL limit is shown as a function of the mass of the lightest chargino and neutralino.

\section{Same-sign dileptons search with b-tagged jets}

With the LHC collision dataset growing, access opens up to rarer production modes, like the electroweak production touched upon in the ATLAS multilepton analysis. At the same time, also more exclusive final states now gain focus, in particular in the context of a light third generation, rendered plausible through arguments of naturelness.

An example of such a search is the same-sign analysis in CMS, but now executed with an additional requirement of two or three b-jets. Asking for two b-jets reduces the already small top background further with a factor of 10. Indeed, because of the selection of two same-sign leptons in top events, the non-prompt isolated lepton will mostly come from one of the two b jets, hence suppressing the probability for that b jet to be identified as such. This background reduction yields further handles to define appropriate search regions as a function of $H_{\rm T}$ and $\met$. For all defined search regions the data was found to be in agreement with the expected background.

\begin{figure}[htb]
  \begin{center}
\raisebox{-0.5\height}{\includegraphics[width=0.48\textwidth]{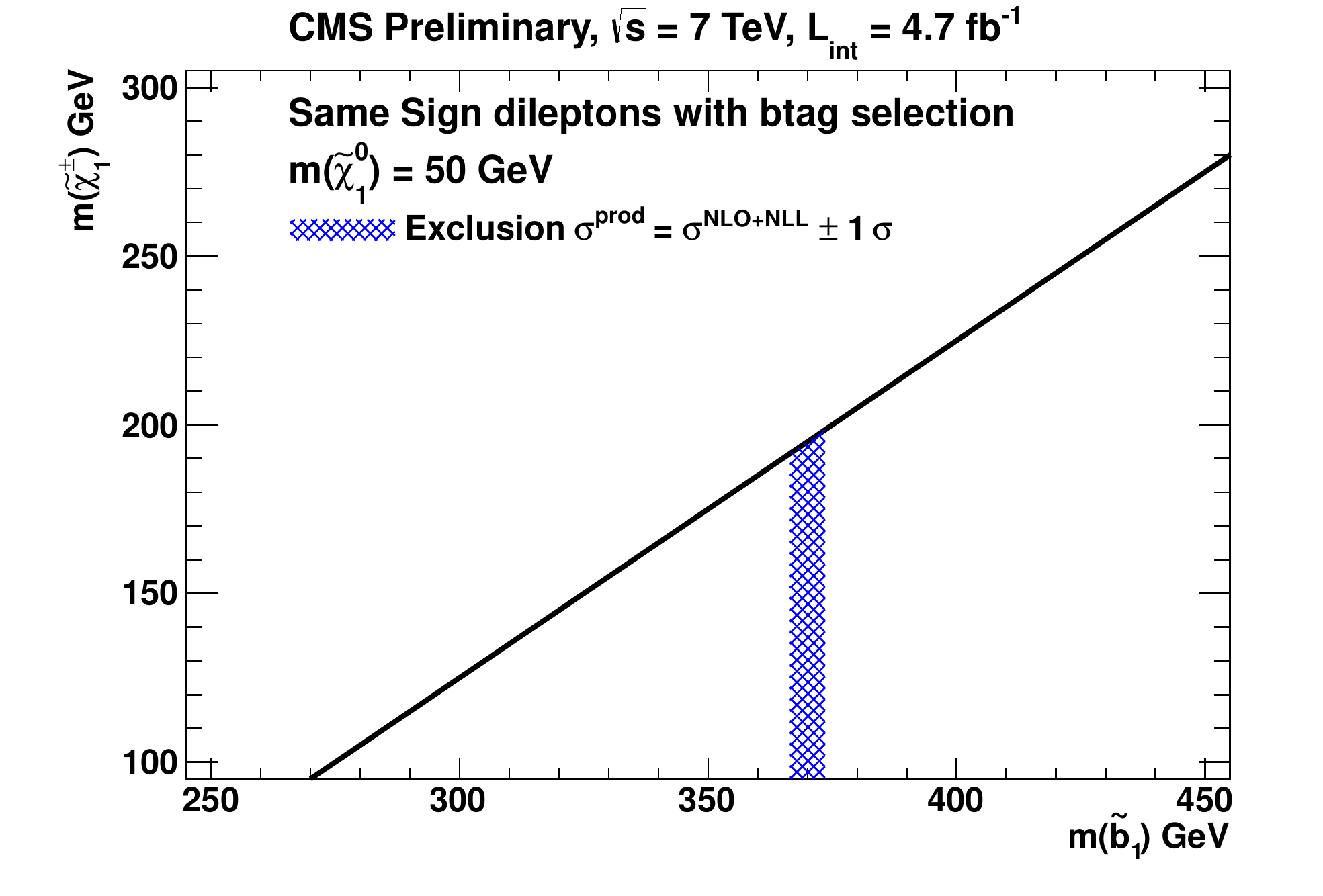}}
    \raisebox{-0.5\height}{\includegraphics[width=0.48\textwidth]{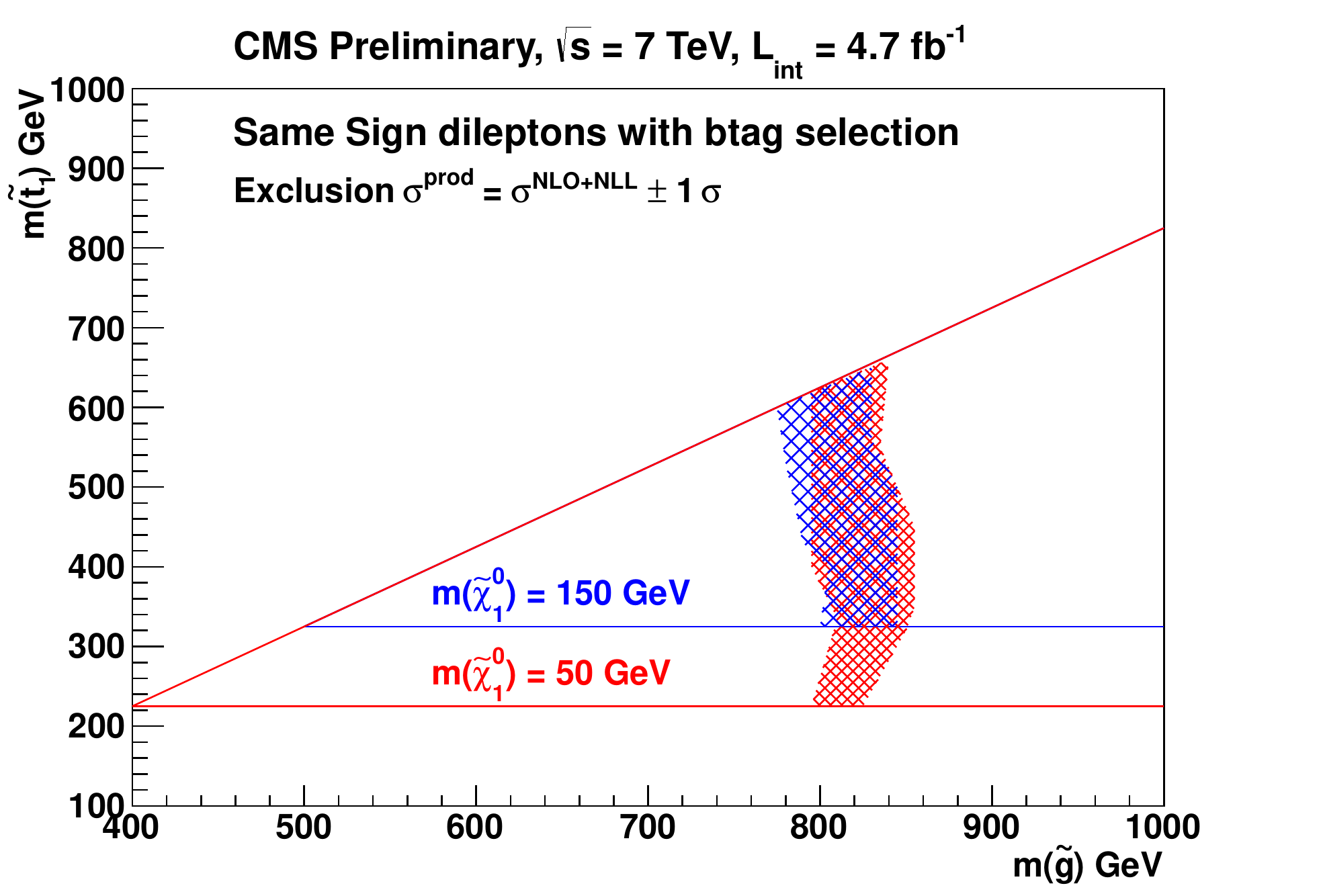}}
      \end{center}
  \caption{Interpretation of the same-sign plus b-jets search in simplified models with direct sbottom production (left) and gluino pair production yielding a four-top final state (right).}
  \label{fig:ssb}
\end{figure}

Two interpretations are of immediate interest for supersymmetry with a light third generation. One considers direct sbottom production, yielding a ${\rm t \bar{t} W^+ W^-}$+ \met{} final state with a decay through an intermediate chargino. In Figure~\ref{fig:ssb} (left) the 95\% CL exclusion limit is shown for this simplified model, as a function of the sbottom and chargino mass, for an LSP mass of $50 \, {\rm GeV}$. It is observed that the mass of the intermediate chargino does not influence the sensitivity of the search. The other interpretation considers gluino pair production, where various decay chains can result in a four-top+ \met{} final state. In Figure~\ref{fig:ssb} (right) the corresponding limit is shown for a simplified model with gluino decay through an intermediate stop, as a function of the gluino and stop masses. Also here it is observed that the sensitivity is driven by the mass of the mother particle, and that the intermediate stop mass plays little role for the sensitivity of the search. In the same vein it was observed that also the exact decay chain leading to the four-top final state only marginally influences the search reach.

\section{Conclusions}

Both the ATLAS and CMS collaborations pursue a vibrant supersymmetry search program. With two orders of magnitude more collision data delivered by the LHC in 2011 compared to 2010, the existing wide variety of searches keeps covering ground rapidly; so far all data has been observed to be compatible with the expected Standard Model backgrounds. As shown with a few examples, search strategies are becoming more advanced, diverse, and targeted, aiming to cover more corners of the SUSY phase space. In particular, dedicated attention is gearing up towards rare electroweak SUSY production, complementing the existing strong production searches, and towards third generation SUSY, which could show up at low masses, favoured by arguments of naturelness. The data currently being collected by the LHC in the 2012 run, now operating at $8 \, {\rm TeV}$ centre-of-mass energy, will provide ample opportunity to continue the quest for new-physics signals of supersymmetry.



\section*{Acknowledgments}
I would like to thank the LHC colleagues for the outstanding performance of the accelerator, and the many members of the ATLAS and CMS collaborations who, from detector operation to analysis, made it possible for so many state-of-the art results to be presented at this conference.


\begin{thebibliography}{99}
\bibitem{susy} References to literature on supersymmetry can be be found in the papers and reports cited below.
\end{thebibliography}
\end{document}